\documentstyle[twoside,fleqn,espcrc2,epsf]{article}
\newcommand{\be}{\begin{equation}}
\newcommand{\ee}{\end{equation}}
\newcommand{\bea}{\begin{eqnarray}}
\newcommand{\eea}{\end{eqnarray}}
\newcommand{\bean}{\begin{eqnarray*}}
\newcommand{\eean}{\end{eqnarray*}}

\newcommand{\AmS}{{\protect\the\textfont2
  A\kern-.1667em\lower.5ex\hbox{M}\kern-.125emS}}

\title{Study of 3-flavor QCD Finite Temperature Phase Transition with Staggered Fermions
\thanks{
This work was supported by the U.S. Department of Energy. The simulations were
conducted on QCDSP machines at  Columbia University and RIKEN-BNL 
Research Center. 
}}

\author{ 
X.~Liao\address{Physics Department, Columbia University, 
New York, NY 10027, USA}}
\begin{document}

\begin{abstract}
We have studied the 3-flavor, finite temperature, QCD
phase transition  with staggred fermions on an $ N_t=4$ lattice. 
By studying a variety of quark masses we have located the critical 
point, $m_c$, where the first order 3-flavor transition ends as 
lying in the region $0.32 \le m_c \le 0.35$ in lattice units.  

\end{abstract}

%
\maketitle
\section{INTRODUCTION}

The flavor and mass dependence of finite temperature QCD phase transition
has been studied extensively in the past \cite{Karsch}. 
Three flavor QCD is especially interesting because of the existence of a second 
order end point of the first order phase
 transition line which is conjectured to belong 
to the 3-d Ising universality class \cite{Gavin}. It is similar to systems such as the
liquid-gas \cite{LiquidVapor}, the 3-d Ising \cite{Ising}, the 3-d 3-state Potts model
\cite{Potts} and the SU(2) Higgs model\cite{ElectroWeak}. It's worth mentioning 
that the field mixing phenomenon \cite{LiquidVapor} is quite important in explaining quantitatively 
some critical behaviors of these systems, but it does not have an important effect on the behaviors we are discussing.

In this study, we extend our previous effort \cite{Columbia90} to locate the 
end point using the standard Wilson gauge action and unimproved staggered fermion action. 
There are many approaches to locate the end point, which is 
charaterized by a diverging correlation length. In this paper, we study the 
discontinuity of the order parameter $\langle \bar{\psi} \psi  \rangle$ \cite{Gottlieb} and the 
meson screening masses (including 
the scalar singlet $\sigma$ meson) \cite{JLQCD} along the first order phase transition line
and extrapolate to the end point. These two approaches works well in the region
with relatively strong first order phase transition. Another method called the Binder 
intersection method \cite{Ising} is based on the 
finite size scaling of symmetry sensitive quantities such as $B_4 = {{\langle 
(\Delta M)^4 \rangle} \over {\langle (\Delta M)^2 \rangle^2}} $, in which M is
the magnetic-like observable. This method can be used to study the region close 
to the end point and does not require a knowledge of critical exponents. However, it's rather 
expensive to obtain a large amount of statistically independent samples from full QCD simulations 
using existing evolution algorithms, especially in the critical region where the
critical slowdown is problematic. The Binder intersection method is currently under 
study and is not reported here.

\section{EVOLUTION}

\begin{figure}[htb]
\vskip -5mm
\hbox{\epsfxsize = 65mm  \epsfysize = 45mm \hskip -1mm \epsffile{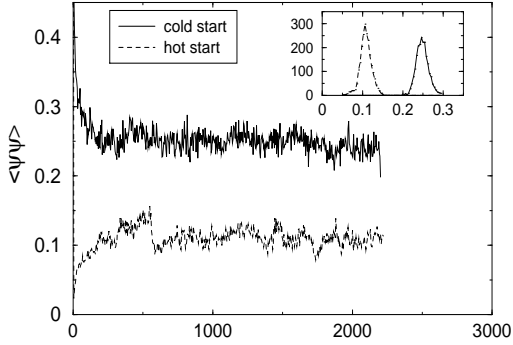}}
\vskip -10mm
\caption{
The $\langle \bar{\psi} \psi \rangle$ evolution and histogram for $m=0.02$ and $L=16$ at $\beta_c=5.1235$
}
\vskip -8mm
\label{fig:pbp_evol_m002_16}
\end{figure}

\begin{figure}[htb]
\vskip -10mm
\hbox{\epsfxsize = 65mm  \epsfysize = 45mm \hskip -1mm \epsffile{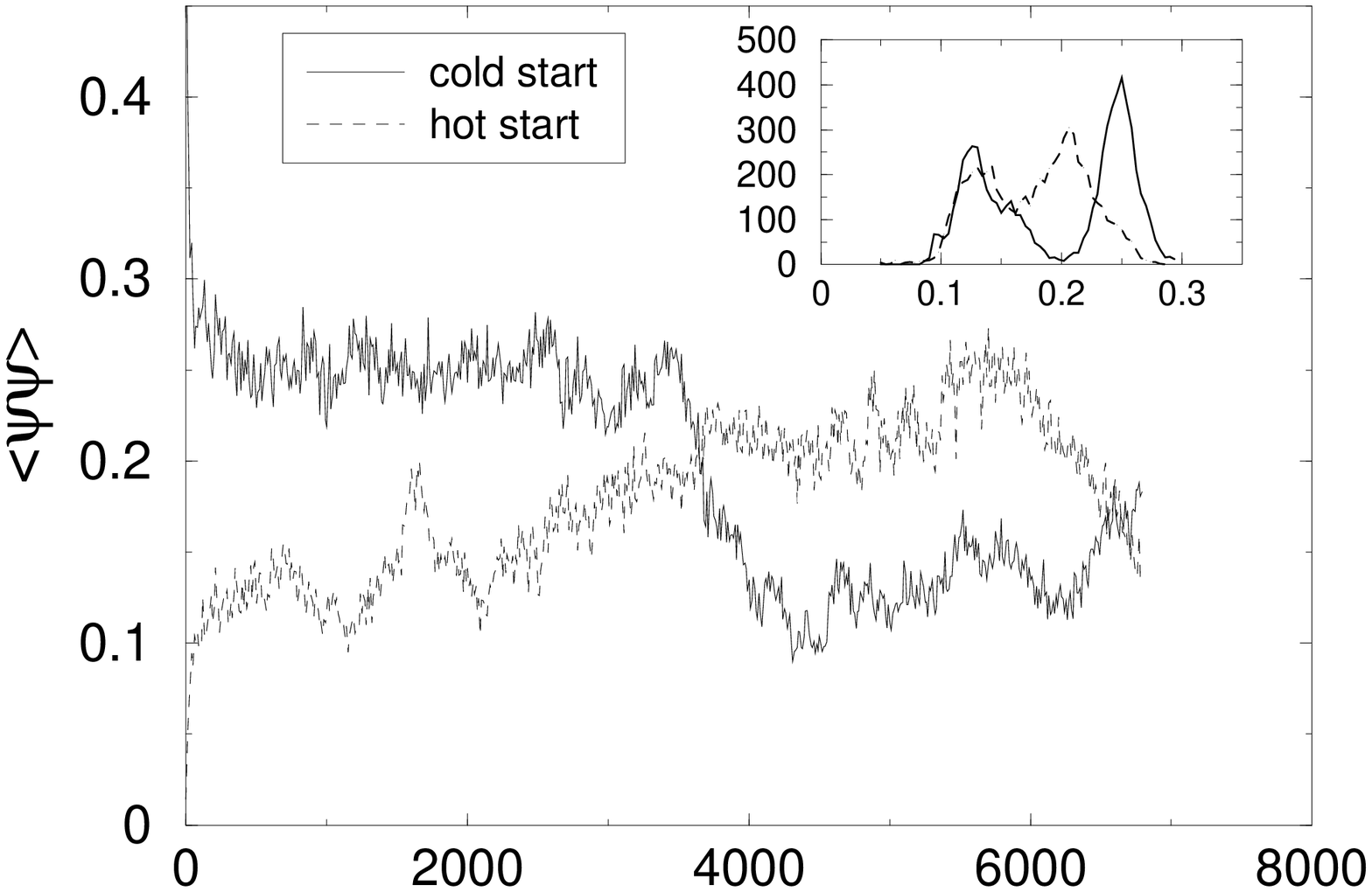}}
\vskip -10mm
\caption{
The $\langle \bar{\psi} \psi \rangle$ evolution and histogram 
for $m=0.025$ and $L=16$ at $\beta_c=5.132$
}
\vskip -8mm
\label{fig:pbp_evol_m0025_16}
\end{figure}

\begin{figure}[htb]
\vskip -2mm
\hbox{\epsfxsize = 65mm  \epsfysize = 47mm \hskip -1mm \epsffile{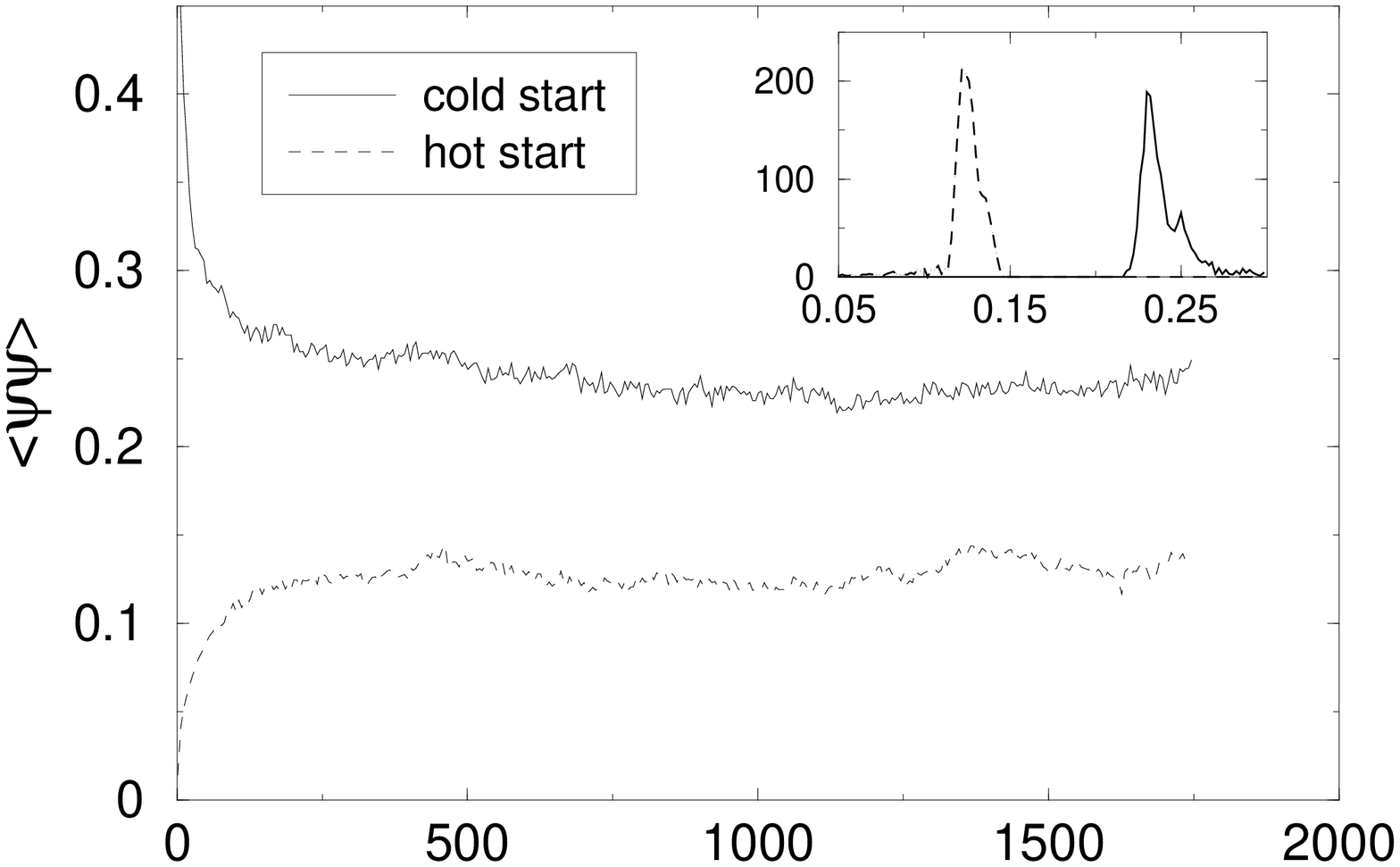}}
\vskip -10mm
\caption{
The $\langle \bar{\psi} \psi \rangle$ evolution and histogram 
for $m=0.025$ and  $L=32$ at $\beta_c=5.132$
}
\vskip -8mm
\label{fig:pbp_evol_m0025_32}
\end{figure}

\begin{figure}[htb]
\vskip -3mm
\hbox{\epsfxsize = 65mm  \epsfysize = 42mm \hskip -1mm \epsffile{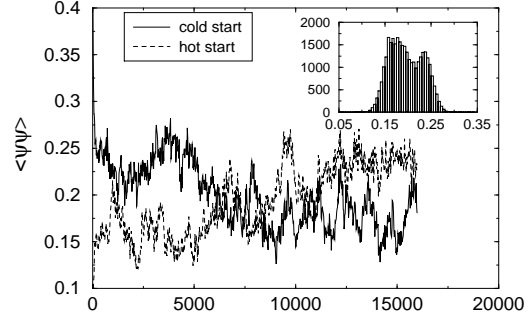}}
\vskip -10mm
\caption{
The $\langle \bar{\psi} \psi \rangle$ evolution and histogram for 
$m=0.035$ and $L=16$ at $\beta_c=5.151$
}
\vskip -8mm
\label{fig:pbp_evol_m0035_16}
\end{figure}

To determine the first order phase transition line, we evolve the system starting from both 
cold and hot initial configurations. We use the hybrid Monte Carlo R algorithm with a
trajectory length of 0.5 molecular dynamic time and step sizes ranging from 0.01 to 
0.0125 depending on quark mass. If a strong first order phase transition exists, we
should be able to observe two-state coexistence at large enough volume. This is 
indeed found at the small quark masses 0.015 and 0.02 
(Fig. \ref{fig:pbp_evol_m002_16})
on an $L=16$ lattice. At $m=0.025$ , tunneling between the two phases happens in 
our simulation
on an $L=16$ lattice (Fig. \ref{fig:pbp_evol_m0025_16}). However, this tunneling is not found on a larger, $L=32$ 
lattice (Fig. \ref{fig:pbp_evol_m0025_32}), indicating that the phase transition is still first order.
For $m=0.035$, no two-state signal is found
on either $L=16$ (Fig. \ref{fig:pbp_evol_m0035_16}) or $L=32$ lattices.  The system also shows significant critical slowdown 
behavior and large fluctuations, which indicate a long correlation length and closeness to a second order 
critical point. Crossover behavior is observed at $m=0.04$.

\section{SCREENING MASS}
\begin{figure}[htb]
\vskip -5mm
\hbox{\epsfxsize = 65mm  \epsfysize = 40mm \hskip -1mm \epsffile{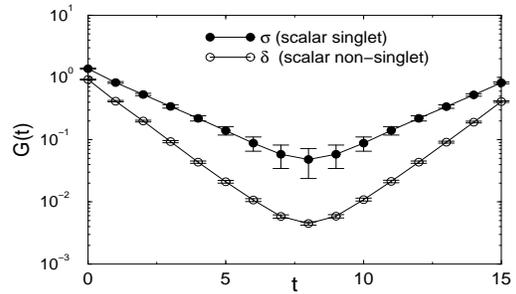}}
\vskip -10mm
\caption{
Screening correlators
}
\vskip -8mm
\label{fig:correlator}
\end{figure}

Theoretical study \cite{Gavin} shows that the scalar singlet meson becomes 
massless while other mesons stay massive at the end point. To improve
statistics, we measure the screening correlators in all three spacial directions 
and average them.  The disconnected diagram needed for the flavor singlet case is measured 
using a noise estimation technique \cite{stagsinglet}. We did a careful study of the effect 
of the type and number of noise sources to ensure that the noise
from noise estimator is smaller than the gauge configuration noise. 
Figure \ref{fig:correlator} shows the time dependence of a typical correlator we obtained.
The meson screening masses are measured at the critical $\beta$ value where a two-state 
signal is observed. It becomes difficult for large quark masses, near $m_c$ which needs 
very large volume to reduce frequency of tunneling events. 

\par
\begin{figure}[htb]
\vskip 0mm
\hbox{\epsfxsize = 70mm  \epsfysize = 50mm \hskip -1mm \epsffile{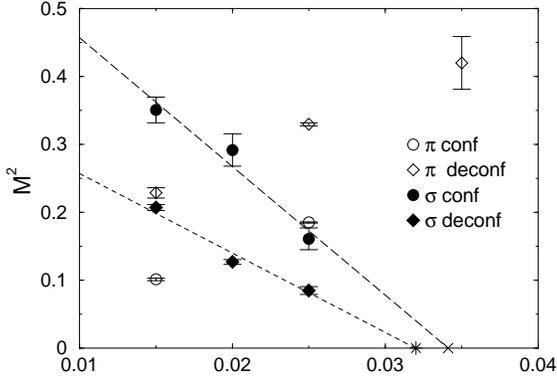}}
\vskip -10mm
\caption{
$(M_{\rm screening})^2$ vs. quark mass.
}
\vskip -8mm
\label{fig:plot_masssq_vs_qkmass}
\end{figure}

Figure \ref{fig:plot_masssq_vs_qkmass} shows the result of the pseudoscalar and scalar flavor singlet ($\sigma$) screening masses.

Extrapolating using the form $M_{\sigma} \propto (m_c - m)^{1/2}$, we get $m_c=0.034(3)$ from the confined
phase and $m_c=0.032(2)$ from the deconfined phase.

\section{ $\langle \bar{\psi} \psi \rangle$ DISCONTINUITY}
\begin{figure}[htb]
\vskip -2mm
\hbox{\epsfxsize = 70mm  \epsfysize = 41mm \hskip -1mm \epsffile{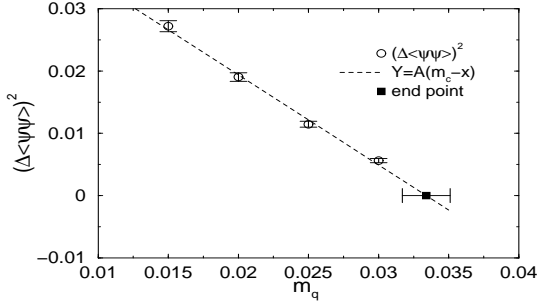}}
\vskip -10mm
\caption{
Discontinuity in $\langle \bar{\psi}\psi \rangle$ with a mean field extrapolation.
}
\vskip -8mm
\label{fig:plot_fit_pbp_diff_sq_vs_mass}
\end{figure}

The discontinuity in the order parameter $\langle \bar{\psi}\psi \rangle$ decreases with
increasing quark mass and finally disappears at $m_c$. We fit the data with the mean field
exponent $\beta=1/2$ (Fig. \ref{fig:plot_fit_pbp_diff_sq_vs_mass}), defined as 
$\Delta \langle \bar{\psi}\psi \rangle \propto (m_c - m)^{\beta}$, and find $m_c=0.0334(17)$. Fitting to the 3-d 
Ising exponent of 0.3285 gives a much poorer $\chi^2$. 

\section{SCALE SETTING}
 We have done a scale setting
calculation on an $8^3\times 32 $ lattice. The pseudoscalar to vector mass ratio is
0.3134(34) and 0.3666(32) for $m=0.025$ and $m=0.035$, respectively. If we use the physical $\rho$ mass
for the vector meson, this implies an SU(3) symmetric pseudoscalar mass of 281 Mev at $m=0.035$.

\section{CONCLUSIONS}

In conclusion, we have obtained the approximate location of the end point by 
studying the behavior of order parameter discontinuity and meson screening 
masses as quark mass approaches the end point of first order phase transition 
line. From our scale setting calculation, we conclude that 
there is no first order phase transition for three flavor QCD at the strange quark mass. 
Higher statistics and larger volume studies in the critical region are needed to 
determine more accurately the critical point, critical exponents and universality 
class.


\begin{thebibliography}{9}
\vskip -0.1cm
\bibitem{Karsch} F. Karsch et al., Nucl.Phys.Proc.Suppl. 94 (2001) 411.
\bibitem{Gottlieb} S. Gottlieb, Nucl. Phys. Proc. Suppl. 20 (1991) 247.
\bibitem{Columbia90} F.R.Brown et al., Phys. Rev. Lett. 65 (1990) 2491.
\bibitem{LiquidVapor} N.B. Wilding, J. Phys.: Condens. Matter 9(1997) 585.
\bibitem{Ising} H.W. Bl\"ote, E. Luijten and J.R. Heringa, J. Phys. A:
Math. Gen. 28 (1995) 6289.
\bibitem{Potts} F. Karsch and S. Stickan, Phys.Lett. B488 (2000) 319-325
\bibitem{ElectroWeak} K. Rummukainen, M. Tsypin, K. Kajantie, M. Laine and
M. Shaposhnikov, Nucl. Phys. B532 (1998) 283.

\bibitem{JLQCD} S. Aoki et al. (JLQCD), Nucl. Phys. B (Proc. Suppl.) 73 (1999) 459.

\bibitem{Gavin} S. Gavin, A. Gocksch and R.D. Pisarski, Phys. Rev. D49 (1994) 3079.

\bibitem{stagsinglet} L. Venkataraman and  G. Kilcup, The Eta-prime Meson with Staggered Fermions,
hep-lat/9711006.
\end{thebibliography}
\end{document}